\documentclass[conference]{IEEEtran}
\IEEEoverridecommandlockouts
\usepackage{cite}
\usepackage{setspace}
\usepackage{amsmath,amssymb,amsfonts}
\usepackage{algorithmic}
\usepackage{graphicx}
\usepackage{subfigure}
\usepackage{subfig}
\usepackage{subfloat}

\usepackage{textcomp}
\usepackage{xcolor}
\usepackage{todonotes}
\def\BibTeX{{\rm B\kern-.05em{\sc i\kern-.025em b}\kern-.08em
    T\kern-.1667em\lower.7ex\hbox{E}\kern-.125emX}}
\begin{document}

\title{Optic Fingerprint: Enhancing Security in Visible Light Communication Networks}

\author{Xuanbang Chen$^{\S,\dagger}$, Ziqi Liu$^{\S}$, Xun Zhang$^{\S}$, Yuhao Wang$^\dagger$, Dayu Shi$^{\S}$, Xiaodong Liu$^\dagger\!\S$,  \\
   $^{\S}$Institut Supérieur D’électronique de Paris, France, $^\dagger$Nanchang University, China \\
Email: \emph{chenxuanbang@email.ncu.edu.cn}, \emph{ziqi.liu@isep.fr}, \emph{xun.zhang@isep.fr}, \\ \emph{wangyuhao@ncu.edu.cn}, \emph{dayu.shi@ext.isep.fr}, \emph{xiaodongliu@whu.edu.cn}
\thanks{This work was supported by a grant of the EU Horizon 2020 program, under the 6G BRAINS project H2020-ICT 101017226. (Corresponding author: Xun ZHANG.)}
\thanks{X. Chen and Z. Liu contributed to the work equally and should be regarded as co-first authors.}
}

\maketitle

\begin{abstract}

In addressing physical layer security issues, hardware fingerprinting has been proven to be a reliable method. Additionally, Visible Light Communication (VLC) technology offers a solution to the spectrum congestion in next-generation wireless communications and is noteworthy for its high security. However, there is currently a lack of a comprehensive and systematic description of the hardware fingerprints and their extraction mechanisms for VLC devices. This study aims to bridge this gap by thoroughly analyzing the hardware fingerprints of VLC devices and proposing an innovative extraction mechanism, thereby enhancing the security and reliability of the physical layer. An Optic Fingerprint (OF) model is proposed based on the LED's inherent circuit characteristics, capable of extracting and processing unique feature vectors with high precision. Through extensive experiments, we demonstrate the model's efficacy, achieving up to 99.3\% accuracy in identifying the same manufactured white LEDs under variable conditions, marking a significant improvement in authentication robustness and interference resistance.


\end{abstract}

\begin{IEEEkeywords}
Visible light communication, physics-based LED circuit model, optic fingerprint, machine learning. 
\end{IEEEkeywords}

\section{Introduction}

\IEEEPARstart{T}{HE} transaction from 5G to 6G networks marks a significant leap in communication technology but raises emergent and complex security challenges \cite{3gpp2018studyfalse,porambage20216g}. The classical cryptographic solutions fall short in these novel challenges due to their complexity-based paradigms. Consequently, innovative security approaches, particularly tailored for 6G systems, are desired from both the industry and academia. Device Fingerprint (DF) is the most potential technology in Physical Layer Security (PLS) enhancing approaches, owing to its unique and secure identification capabilities.  It leverages the inherent hardware characteristics of devices, guaranteeing non-replicability and distinctive identification \cite{liyanage2016opportunities, khan2019survey}, offers a more robust and adaptable security framework, well-suited to meet the demands of 6G and IoT applications \cite{bloch2011physical,aazhang2019keydrivers,ziegler2021security,khorsandi2022hexa,ieee2022security}. 

Most of the existing research on DF focuses on Radio Frequency (RF) application scenarios. Researchers proposed the concept of Radio Frequency Fingerprint (RFF) to achieve the 92.29\% accuracy of base station authentication for Wi-Fi, LTE, ZigBee, and etc. \cite{xie2024radio,zhang2023radio,xu2015device,sun2020construction,yu2016rf,merchant2018deep}. Whereas in the future 6G era, the available spectrum extended to the optical band promoting heterogeneous physical layer technologies\cite{jiang2021road}. 
Visible Light Communication (VLC) \cite{chi2020visible}, leveraging signal spectrum from 380nm to 780nm to transmit signal through the optical wireless channel, possesses attributes like high capacity, ultra-high data rates, low latency, and inherent security due to the limited penetration of optical signals. 


However, few researches address the issues of DF in the VLC domain. Existing studies utilize frequency response measurements, notably the S21 parameter, to develop a DF model\cite{shi2020device}\cite{shi2021improving} and employ machine learning or deep learning approaches\cite{liu2022alexnet} to detect DF features, aiming to authenticate device identities effectively. These studies, however, are largely limited to static environments, overlooking the impact of environmental and spatial fluctuations, such as distance, angle, and signal noise interference. Such neglect results in compromised anti-noise and spatial stability, thereby undermining the adaptability of these models in complex environments.

This paper proposes a new DF model in the VLC system, which is named the Optic Fingerprint (OF) model, and the corresponding machine learning-based feature extraction mechanism. Significant contributions of this study include:
\begin{itemize}
    \item  An OF model is proposed to characterize the unique nonlinearity attributes of LEDs to form a reliable feature vector.
    \item The extraction scheme for OF and the security identification framework is developed, utilizing power spectrum analysis and a machine learning classifier for precise device identification against an authorized fingerprint database.
    \item Numerous experiments are performed to validate our proposed OF model. The proposed OF model shows a high accuracy of 99.3\% in identifying commercial white LEDs. Compared to the traditional S21 fingerprint, the proposed OF model presents a lower complexity with better performance of anti-environmental interference.

\end{itemize}

The rest of this paper is organized as follows. Section II outlines the OF model and extraction methodology. Section III details the feature extraction and verification process. Section IV presents the identification accuracy of the proposed fingerprint. Section V concludes the paper, discussing the implications of our future directions.

\section{Physics-based Optic Fingerprint Model}

\begin{figure}[th!]
\centerline{\includegraphics[width=2.9in]{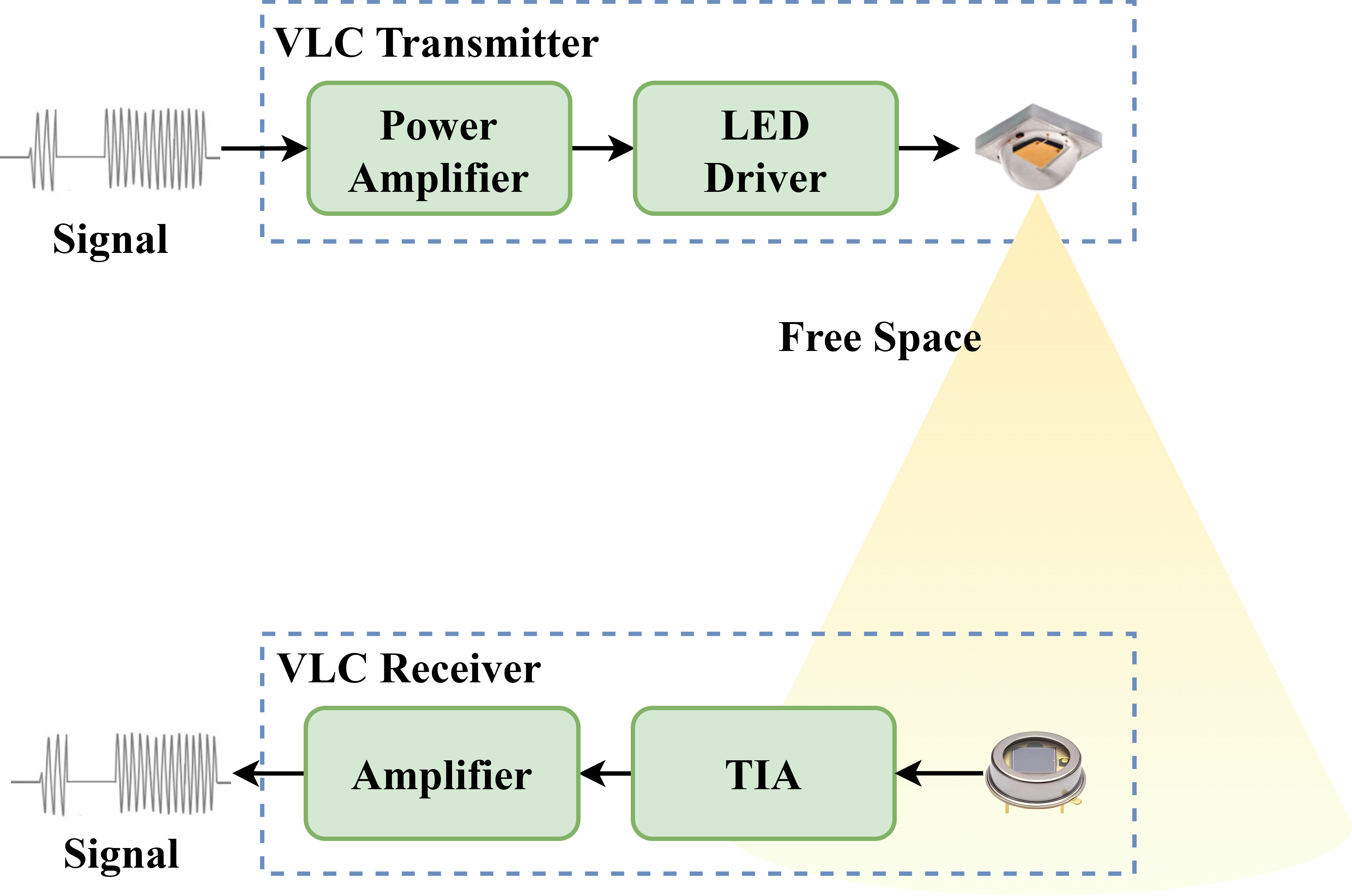}}
\caption{The typical VLC link diagram.}
\label{VLC_block}
\end{figure}

A novel OF model is proposed based on the physics variations of LED devices for security enhancement in this context. Generally, as shown in Fig. \ref{VLC_block}, an LED-based VLC system comprises a VLC transmitter, Free Space, and VLC receiver. Specifically, the power amplifier amplifies the signal to drive LEDs with lower AC impedance, while the LED driver converts the voltage signal into a current signal, achieving amplitude modulation of the LED luminous intensity. The optical modulated signal emitted by the VLC transmitter travels through the channel and is intensity-detected by the VLC receiver. It is noteworthy that despite belonging to the same batch, the hardware components of the VLC transmitter, including the power amplifier, driver, and LED, may exhibit slight variations due to manufacturing tolerances. These hardware differences influence the optical signal, which is subsequently transmitted to the receiver. The received optical signal, denoted as $y(t)$, can be expressed as
\begin{equation}
    \begin{aligned}
    y(t) = G_{\rm PA}G_{\rm D}H_{\rm C}G_{\rm Re}\int_{-\infty}^{\infty} x(t)h_{\rm LED}(t-\tau)d\tau,
    \end{aligned}
    \label{receiver-signal}
\end{equation}

$x(t)$ denotes the electrical signal source. $G_{\rm PA}$, $G_{\rm D}$ represent the gain of the power amplifier and driver, respectively. Moreover, $h_{\rm LED}(t)$ is the impulse response of the LED. It is worth noting that the LED, as a bandwidth-limited communication device, is the main source of hardware nonlinearity and variance. Since the non-line of sight (NLoS) component of the VLC channel is very weak, the channel response $H_{\rm C}$ is considered to be a constant loss when only considering the line of sight (LoS) component, which can be calculated as
\begin{equation}
    \begin{aligned}
    H_{\rm C} = \frac{(m+1)A_{\rm r}}{2\pi d^2}\cos^m(\phi)g(\psi)\cos(\psi),
    \end{aligned}
    \label{lambert}
\end{equation}

where $m=-\rm{ln2/ln(cos\phi_{1/2})}$ is the Lambertian emission order and $\phi_{1/2}$ is the emission semi-angle of LED. $A_{\rm r}$ is the physical detection area of the receiver front-end. The channel loss is related to the distance $d$ (between the LED and the receiver front-end) and the irradiance angle $\phi$ of the light \cite{channel}. 

Thus, amplifier, driver, and channel impulse responses can all be considered linear parameters. Therefore, modeling the nonlinearity of LEDs is key to characterizing the distinctive feature of each device. Numerous generic mathematical modeling methods have been proposed to characterize the modulation nonlinearity of LEDs, such as the memory polynomial model, Volterra model, and Hammerstein model \cite{dengmodel}. However, these modeling methods typically exhibit high implementation complexity and fail to capture the inherent common modulation nonlinearity of LEDs, which are strongly correlated with the physical structure of the LED. Our previous work modeled the GaN LED from the physics aspect, through the analysis of carrier concentrations in each LED layer, the carrier diffusion, capture, thermal escape, and recombination are formulated by carrier rate equations \cite{dayu}. Based on the previous work, the physics-based LED model is improved to completely characterize the high-frequency characteristics in this paper, and a novel OF model is proposed.

\begin{figure}[htbp]
\centering
\subfigure[]{
\includegraphics[width=3.3in]{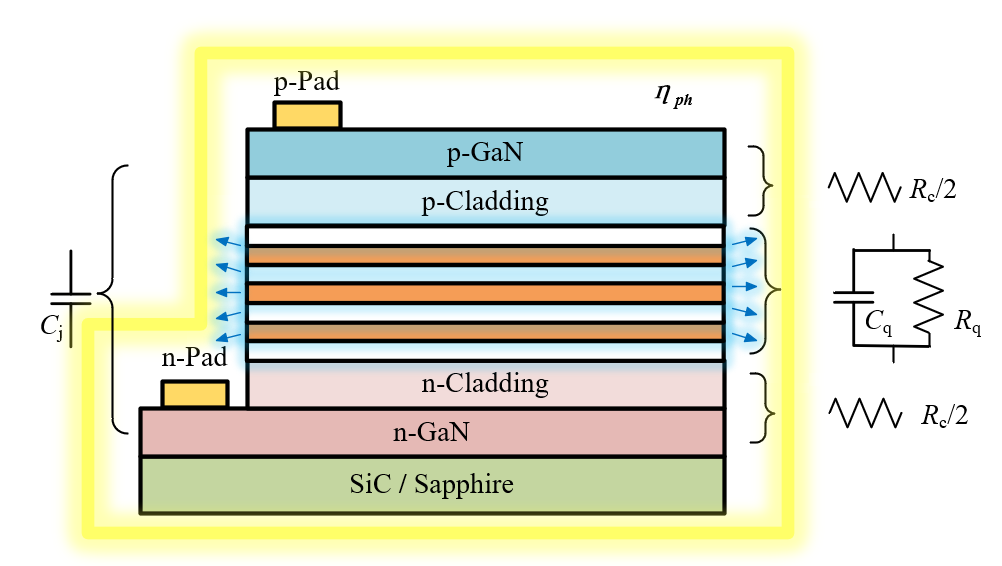} 
}
\subfigure[]{
\includegraphics[width=3.3in]{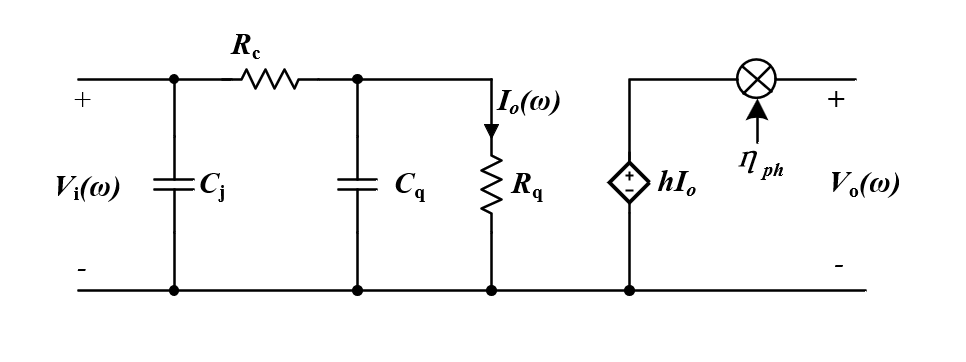} 
}
\caption{(a) The physical structure and (b) the equivalent circuit model of the MQW LED.}
\label{LED-model}
\end{figure}
As shown in Fig. \ref{LED-model}(a), silicon carbide (SiC) and sapphire are the common substrates used in the LED. Due to the MQW being wrapped by a cladding layer, it is necessary to consider the capacitance and resistance effects of the quantum well in the LED model. Thus, $R_{\rm q}$ and $C_{\rm q}$ represent the equivalent resistance and capacitance of MQW, respectively. Specifically, the current through $R_{\rm q}$ represents the recombination of electrons, and $C_{\rm q}$ reflects the charge storage effect in the quantum well \cite{rashidi2017differential}. Moreover, $ R_{\rm c}$ is the equivalent resistance distributed in p- and n-Cladding layers. $C_{\rm j}$ is the sum of the barrier capacitance and the parasitic capacitance. Thus, the corresponding equivalent circuit model is schematically illustrated in Fig. \ref{LED-model}(b). It is worth noting that $h$ represents the external quantum efficiency. $\eta_{ph}$ is the response of phosphor. The blue wavelength photons emitted by the LED are excited by the phosphor to produce yellow light and are mixed into white light to provide white lighting services to users. The analytical expression of the LED’s impedance $Z_{\rm LED}(w)$ can be written as follows. 
\begin{equation}
\begin{aligned}
& Z_{\rm LED}(w) = \frac{1}{jwC_{\rm j}} //(R_{\rm c}+R_{\rm q} //\frac{1}{jwC_{\rm q}})  \\
&= \frac{R_c+R_q+jwR_cR_qC_q}{1-w^2R_cR_qC_qC_j+jw(R_qC_q+R_cC_j+R_qC_j)},
\end{aligned}
\label{F LED impedance}
\end{equation}

where $w$ is the angular frequency and $j$ is the imaginary unit. The operator $//$ denotes the parallel calculation in the circuit. Thus, the parameters of the LED equivalent model can be extracted from the measured impedance curve. However, this method is less practical, as it necessitates additional and complex experiments to obtain device impedance features. In this context, the frequency response of the LED is exploited to extract model features, allowing the utilization of existing communication data for obtaining the transmission response without additional experiments. Specifically, the transfer function of the LED-based VLC system can be expressed as
\begin{equation}
\begin{aligned}
     &H_{\rm VLC}(w)=G_{\rm PA}G_{\rm D}H_{\rm C}G_{\rm Re}H_{\rm LED}(w) \\ 
     &=\frac{G_{\rm PA}G_{\rm D}H_{\rm C}G_{\rm Re}h}{1-w^2R_cR_qC_qC_j+jw(R_qC_q+R_cC_j+R_qC_j)}.
     \label{H_VLC_FP}
\end{aligned}
\end{equation}
As $G_{\rm PA}$, $G_{\rm D}$, $H_{\rm C}$, $G_{\rm Re}$, and $h$ remain constant at a specific measurement point, a uniform $\zeta$ is employed to represent these parameters, simplifying the transmission formulation. Therefore, the feature parameters of the proposed LED model are summarized as a five-dimensional vector $R_c$, $C_j$, $R_q$, $C_q$, and $\zeta$, which can be derived from the VLC system response $H_{\rm VLC}$. Additionally, $H_{\rm C}$ excludes the phosphor response component, as the receiver typically employs a blue light filter to eliminate signal transmission delays. Therefore, the feature parameters $R_c$, $C_j$, $R_q$, $C_q$, and $\zeta$ of the proposed LED model can be derived from the VLC system response $H_{\rm VLC}(w)$. 

\begin{equation}
    \begin{aligned}
    \lambda = [R_c,C_j,R_q].
    \end{aligned}
    \label{receiver-signal}
\end{equation}

The values of $R_c$, $C_j$, and $R_q$ form the OF model represented by the three-dimensional feature vector $\lambda$. It is important to note that the factors $\zeta$ and $C_q$ are omitted from the OF model. This exclusion is justified by the fact that $\zeta$ primarily relies on the testing equipment types and specific positions, rather than the intrinsic characteristics of the LED. Additionally, Section IV provides further evidence that $C_q$ lacks distinct characteristics among different LED devices. $\lambda$ depends on the inherent non-linearity characteristic of each LED, which indicates its unique fingerprints.

\section{Extraction and Identification}
\subsection{Optic Fingerprint Extraction}
\begin{figure}[th!]
\centerline{\includegraphics[width=3.5in]{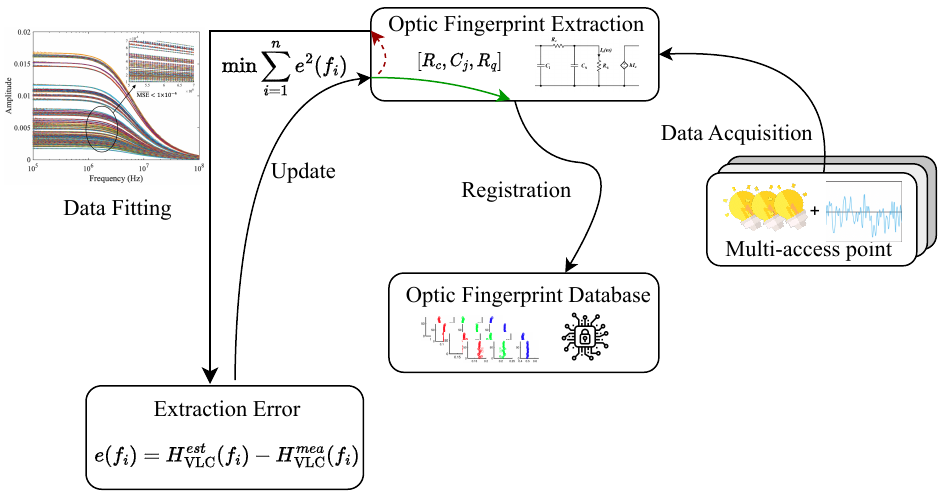}}
\caption{The extraction scheme of the proposed OF model.}
\label{OF_extraction}
\end{figure}

Fig. \ref{OF_extraction} illustrates the OF feature extraction process. The LED is the device for the proposed OF feature extraction, which acts as the access point within the VLC system. The test signal, encompassing options such as a swept, baseband, or modulated signal, undergoes LED-induced nonlinearity and is captured by the VLC receiver. The system response data S21, derived from processing the received signal, is utilized for parameter extraction in the proposed OF model. The extraction algorithm employs the nonlinear least squares method. Specifically, given an initial fingerprint value, the fingerprint feature value continuously updates to minimize the residual $\sum_{i=1}^{n}e^{2}(f_i)$ between the fitted system response data and the measured data. Note that $n$ denotes the frequency points of S21 data. The optimal extraction result is then recorded as the LED's fingerprint feature. Subsequently, numerous LEDs are incorporated into the OF database after feature extraction, facilitating device security authentication.

\subsection{Security Identification Scheme}
\begin{figure}[th!]
\centerline{\includegraphics[scale=0.36]{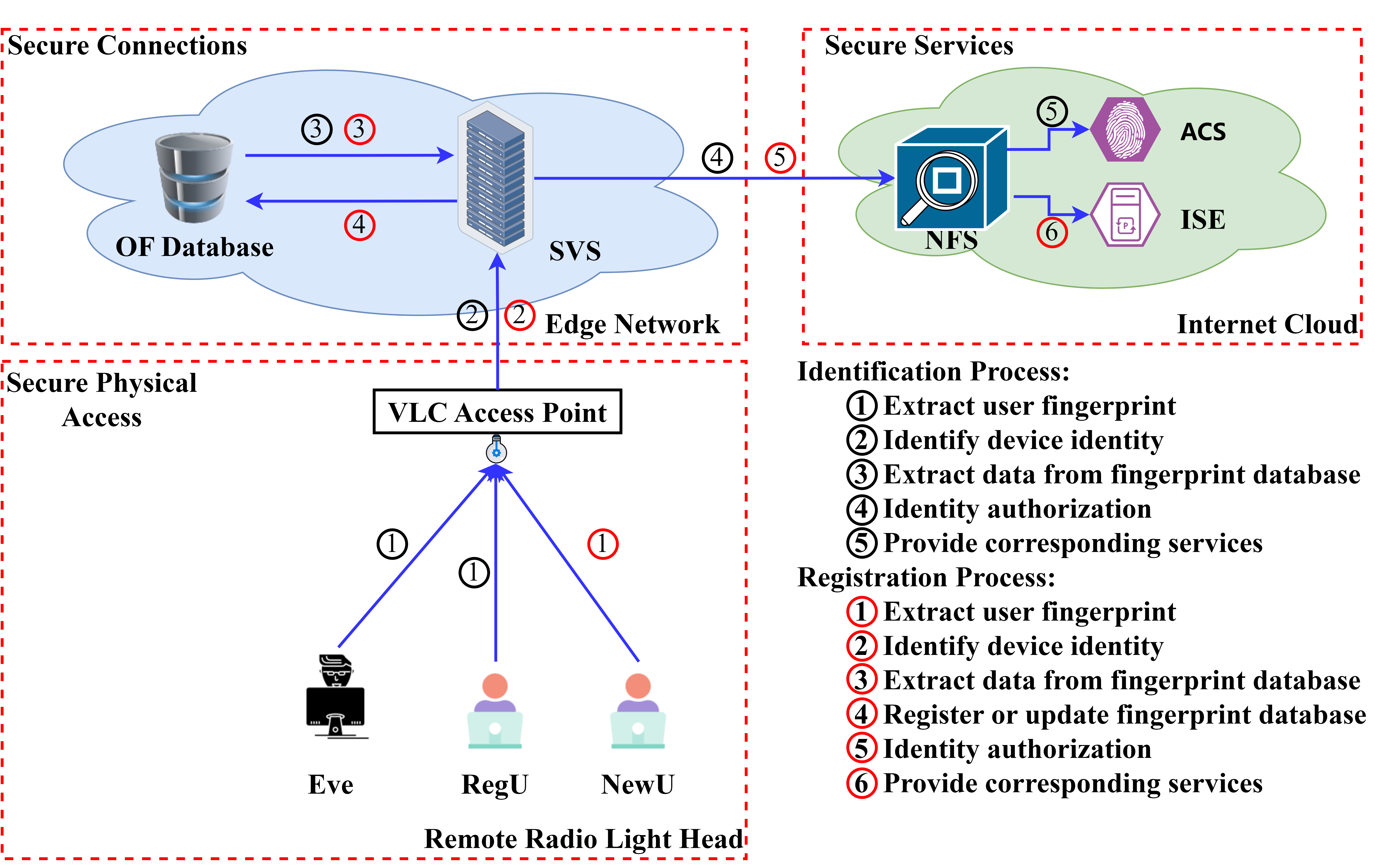}}
\caption{Security identification scheme based on OF model.}
\label{fig2}
\end{figure}

Fig. \ref{fig2} presents an intricate security identification architecture predicated on our OF identification methodology. This system is engineered to preemptively negate network security breaches perpetrated by eavesdroppers (Eve) through identity falsification and unauthorized data acquisition. In the edge network, OF data for registered devices are meticulously cataloged within the OF database. The Security Validation Server (SVS) engages in conjunction with the VLC Access Point to periodically solicit OF data from user equipment (UE). The SVS rigorously evaluates the OF against the established registry of the authenticated user (RegU) fingerprints within the database, subsequently conveying the verification results to the Network Management System (NMS) housed within the Internet Cloud to complete the security validation process. The NMS, predicated on the analysis, orchestrates access authorization via the Access Control Server (ACS) or initiates preemptive alerts through the Identity Services Engine (ISE). The detection mechanism's protocol unfolds as follows:

\begin{enumerate}
\item The VLC Access Point periodically retrieves OF data from UEs.
\item The extracted data are then transmitted to the SVS within the edge network.
\item The SVS concurrently retrieves and analyzes corresponding OF data from the database.
\item The SVS's verdicts are dispatched to the NMS to facilitate security services.
\item The NMS adjudicates network access or restricts it through the ACS based on the analysis.
\end{enumerate}

Further, the framework accommodates the registration of new users (NewU), integrating them seamlessly into the network. The registration process entails:

\begin{enumerate}
\item Upon a NewU's registration request, the VLC Access Point captures pertinent user data and transmutes it into OF data.
\item This data is relayed to the SVS for processing.
\item The SVS concurrently retrieves and analyzes corresponding OF data from the database.
\item The SVS updates the OF database with new entries to the OF database
\item The SVS concurrently communicates with the NMS.
\item The NMS, leveraging the ISE, disseminates new registration details and provisions services accordingly.
\end{enumerate}

This meticulously orchestrated sequence of operations fortifies the network's defense mechanisms against illicit access while contemporaneously ensuring the OF database remains current, thus fostering a secure and adaptable network milieu for both extant and nascent users.

\section{Experiment Setup and Results}
\subsection{Experiment Setup} 
\begin{figure}[th!]
\centerline{\includegraphics[width=3.4in]{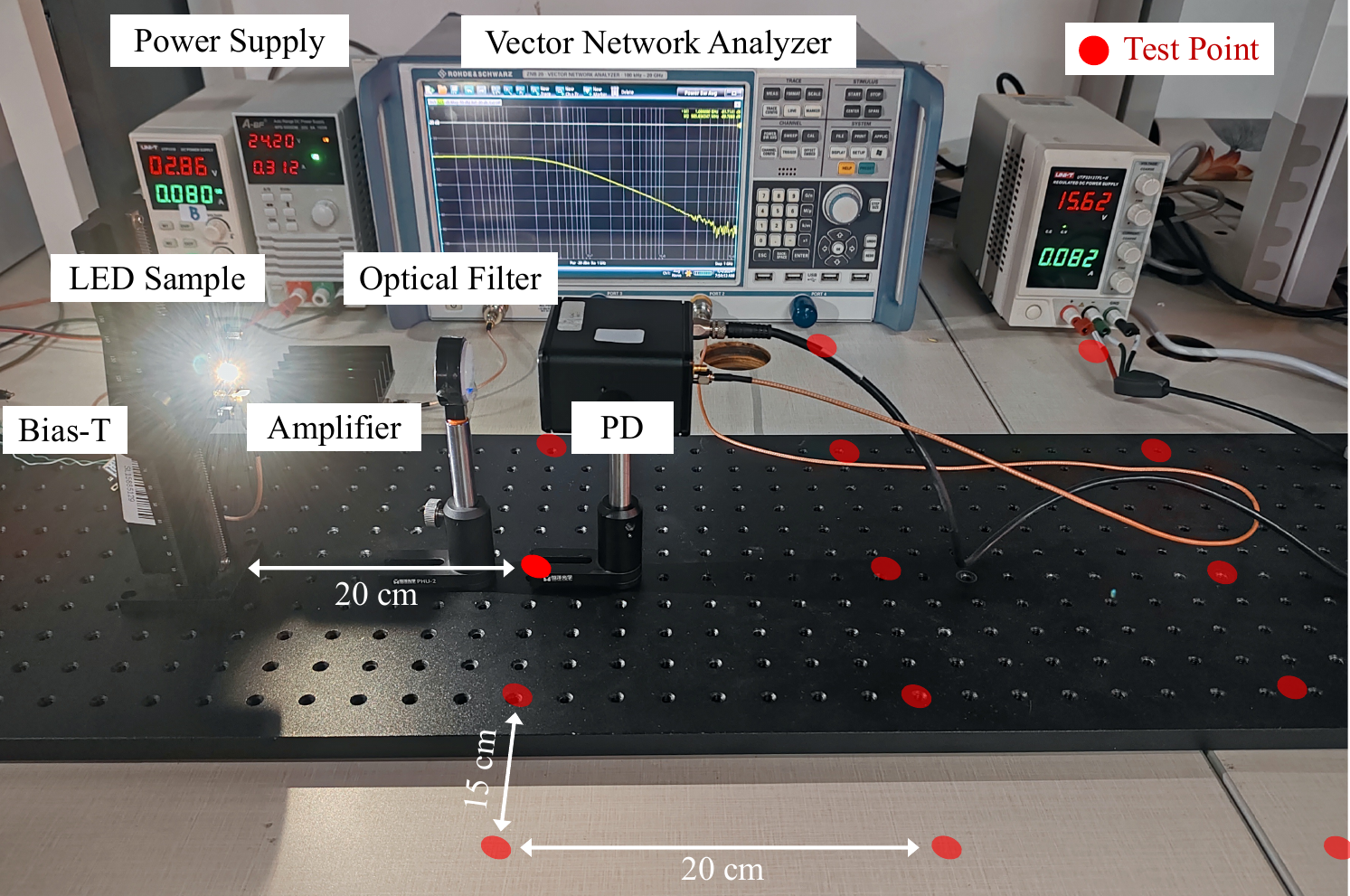}}
\caption{Schematic diagram of the experiment setup.}
\label{fig4}
\end{figure}

The experiment test-bed for acquiring the OF is shown in Fig. \ref{fig4}. Specifically, the vector network analyzer (VNA, Rohde$\&$Schwarz, ZNB20) is employed to measure the frequency responses of the LED. The measurements utilize a continuous wave (CW) frequency-sweep technique. Four LED samples (Cree, XPE2-White) are exploited to establish the OF database. Here, the VNA injects a CW signal with an electrical power of -5 dBm into the power amplifier (Mini-circuits, ZHL-6A-S+), sweeping the modulation frequency from 100 KHz to 100 MHz. Bias-T (Mini-Circuits, ZX85-12G-S+) is exploited to drive the LED with the amplified signal. The VNA then assesses the amplitude and phase variations between the transmitted and received signals, providing the frequency response of the test setup, which includes the wireless link and the optical front ends. The measurements are taken at 15 distinct positions, with each position measured 10 times, totaling 150 measurements per LED. It is worth noting that the longest test distance is 60 cm due to the limited power of a single LED. These measurements form the basis for extracting the normalized feature vectors of each LED. 

Moreover, for each LED, 50\% of the dataset was randomly allocated as the training set, with the feature vectors serving as signatures for authenticated devices. A range of machine learning algorithms—Fine Tree (F-Tree), Kernel Naive Bayes (KNB), Quadratic Support Vector Machine (Q-SVM), Fine K Nearest Neighbor (F-KNN), Ensemble Bagged Trees (EB-Trees), Ensemble Subspace K Nearest Neighbor (ES-KNN), and Narrow Neural Network (NN-Network)—was applied to stratify the training data into four clusters. Following this registration, the remaining data constituted the test set. Alongside, the original, unprocessed S21 parameter data were subjected to the same machine learning classification to provide a baseline for performance comparison.

To rigorously evaluate the identification process, noise disturbances were superimposed onto the test dataset and the unprocessed S21 data. These disturbances ranged from 0 dBm to 90 dBm in 5 dBm increments, resulting in a total of 2550 sets of data, emulating various channel conditions, and testing the robustness of the machine learning models. The analysis compared the performance of the models on both the noise-augmented dataset and the original S21 data, highlighting the efficacy of the OF features in maintaining high identification accuracy despite the presence of environmental noise.

\subsection{Results and Analysis}
\begin{figure}[th!]
\centerline{\includegraphics[width=3.2in]{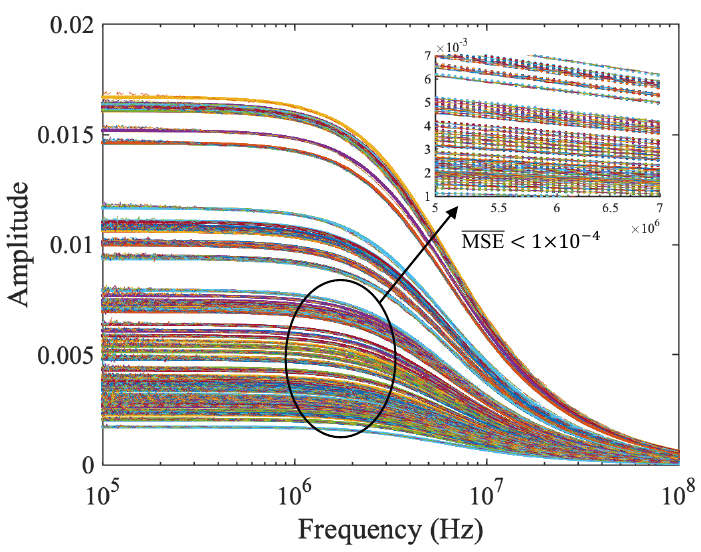}}
\caption{The fitting results of four LEDs in different positions.}
\label{S21}
\end{figure}

The training OF database comprises 300 samples from 4 LED samples. OF extraction is accomplished by fitting S21 to the objective function Eq. (\ref{H_VLC_FP}). Fig. \ref{S21} illustrates the fitting results, compared to 300 S21 measured samples. The frequency response of four LEDs shows different amplitudes due to the various test locations.
Moreover, it can be seen from the subfigure in Fig. \ref{S21} that fitted dots show good coincidence with the measured lines, yielding an average Mean Squared Error (MSE) of less than 1e-4.

\begin{figure}[htbp]
\centering
\subfigure[LED1]{
\includegraphics[width=3.3in]{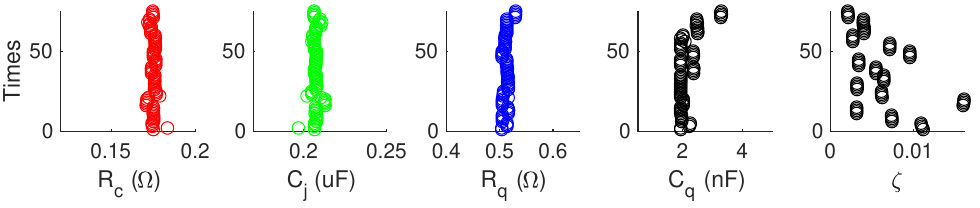} 
}
\subfigure[LED2]{
\includegraphics[width=3.3in]{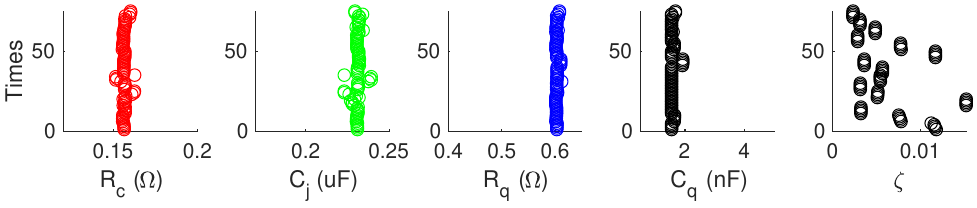} 
}
\subfigure[LED3]{
\includegraphics[width=3.3in]{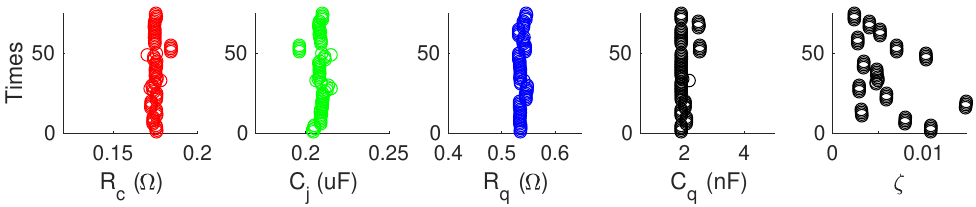} 
}
\subfigure[LED4]{
\includegraphics[width=3.3in]{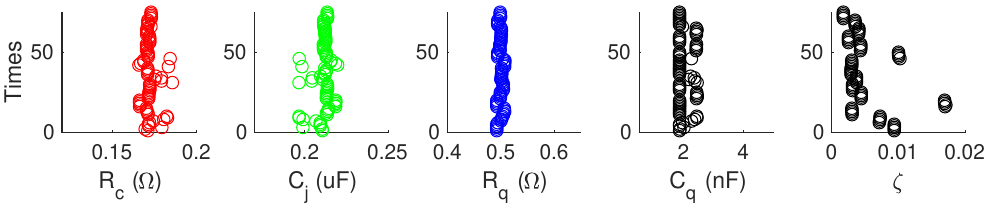} 
}
\caption{The extracted features of four LED samples.}
\label{5FP}
\end{figure}

\begin{figure}[th!]
\centerline{
\includegraphics[width=3in]{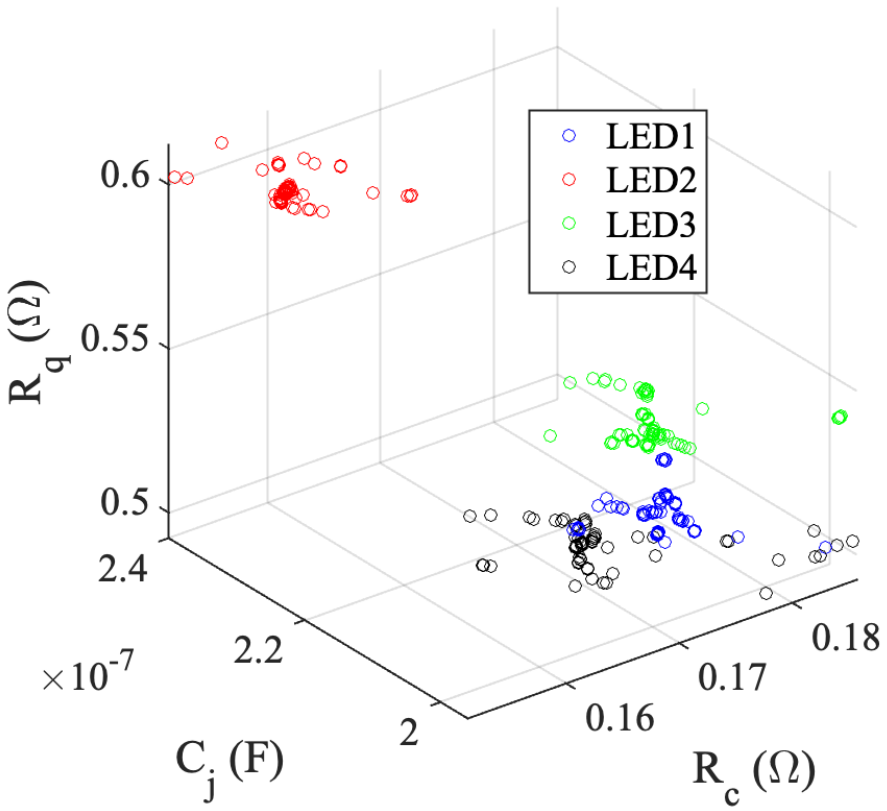}}
\caption{The clustering of the proposed OF model [$R_c$, $C_j$, $R_q$].}
\label{FP_cluster}
\end{figure}

Furthermore, Fig. \ref{5FP} presents the five extracted parameters of the equivalent circuit model based on the aforementioned curve fitting. It can be seen that the circuit model parameters $C_q$ and $\zeta$ of four LED samples are difficult to distinguish. This is because the impact of different test points on $\zeta$ blurs its individual, and $C_q$ is in the nF range making its characteristics challenging to distinguish between LEDs. Thus, the feature vectors [$R_c$, $C_j$, $R_q$] are chosen as the proposed OF model. Fig. \ref{FP_cluster} shows the clustering result of the proposed OF for four LED samples, which are clearly clustering at different locations in the space. 

\begin{figure}[th!]
\centerline{\includegraphics[width=3.2in]{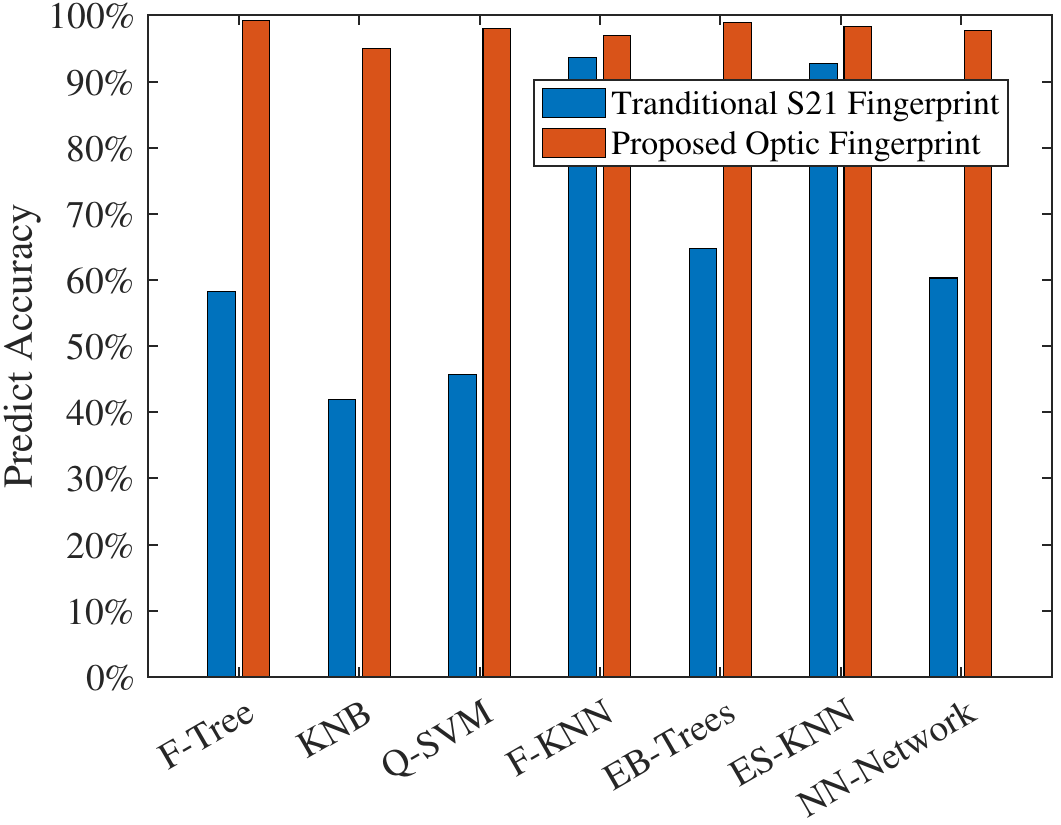}}
\caption{The accuracy comparison of different algorithms using traditional S21 fingerprint and OF models.}
\label{LED_fingerprint_accuracy}
\end{figure}

In the subsequent analysis, various machine-learning algorithms are employed to assess the classification accuracies and validate the efficacy of the proposed OF model. As depicted in Fig. \ref{LED_fingerprint_accuracy}, the OF model consistently achieves remarkable accuracy ranging from 99.3\% to 95.0\% with different algorithms. In contrast, the S21 fingerprint model \cite{shi2020device}, under the same machine-learning evaluations, exhibits a broader spectrum of accuracy rates, peaking at 93.7\% and dropping to as low as 42.0\%. Notably, the OF model requires only three features, whereas the S21 fingerprint model relies on 750 features. This substantial reduction in feature count, combined with superior accuracy and stability, highlights the advantages of our proposed OF model over the conventional S21 model, offering improved efficiency and reduced data complexity.

\begin{figure}[th!]
\centerline{\includegraphics[width=3.4in]{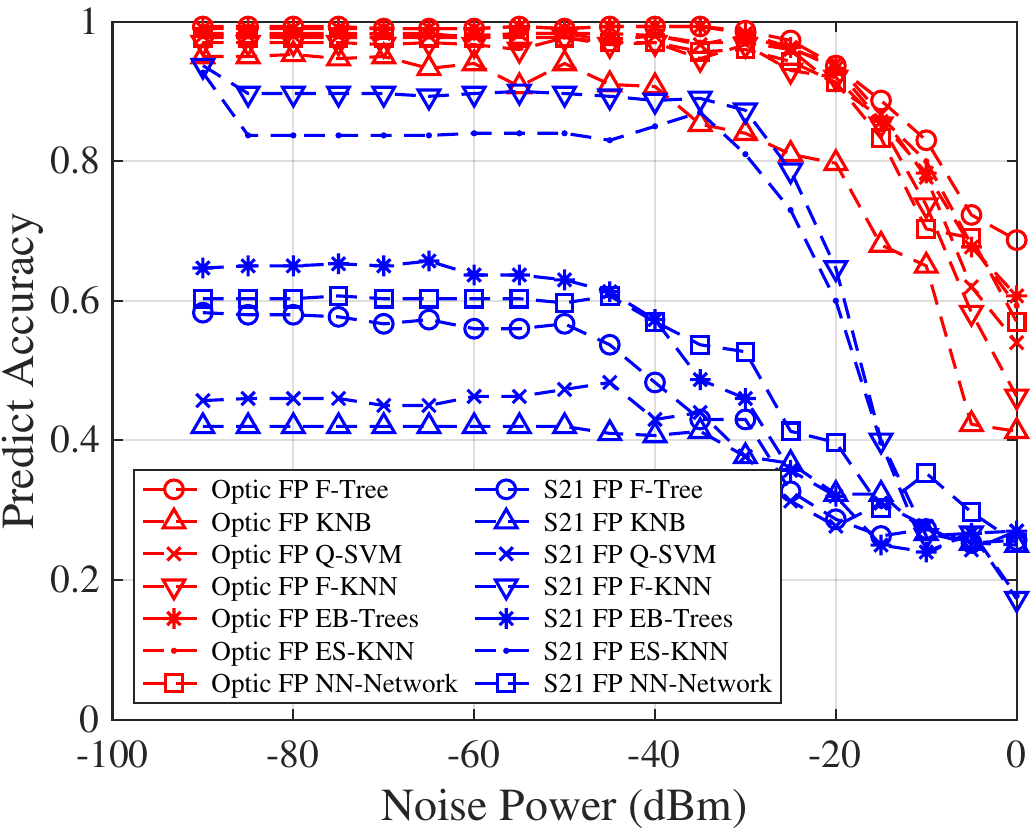}}
\caption{The comparison of anti-noise performance between traditional S21 fingerprint and proposed OF models.}
\label{LED_fingerprint_anti}
\end{figure}

Finally, the reliability of the OF model is analyzed by adding Gaussian white noise of varying powers to both the OF and the traditional S21 fingerprint models. Fig. \ref{LED_fingerprint_anti} illustrates that the proposed OF model consistently achieves accuracies exceeding 80\% across a noise power range from -90 dBm to -20 dBm. In contrast, the S21 fingerprint model demonstrates inferior noise immunity, with accuracy declining to below 62\%. Furthermore, while the performance of the OF model begins to degrade for noise levels surpassing -20 dBm, it remains superior to the S21 fingerprint model. This indicates that fingerprint feature vectors extracted from LED models inherently capture the LED's characteristics, exhibiting robustness against changes in external channel conditions.

\section{Conclusion}


This paper introduces a novel Optic Fingerprint (OF) to apply for enhancing the physical layer security of the sixth-generation network.
Four LED samples were exploited to verify the mechanism of the proposed OF model. High classification accuracies were achieved based on the generic machine learning algorithms, which up to 99.3\%. Moreover, the OF model was compared with the traditional S21 fingerprint model, showcasing a lower complexity with better performance of anti-environmental interference. It is worth noting that the LED samples with more different types and manufacturers will be discussed in future work. The effects introduced by the different receivers will be considered as well, ensuring its effectiveness in the sixth-generation security framework.

\bibliographystyle{IEEEtran}
\bibliography{reference}

\end{document}